\documentclass{PoS}

\usepackage{amsmath,amsfonts,amssymb, graphicx}
\usepackage{subfigure}

\newcommand{\be}{\begin{equation}}
\newcommand{\ee}{\end{equation}}
\newcommand{\bea}{\begin{eqnarray}}
\newcommand{\eea}{\end{eqnarray}}

\title{\vspace*{-1cm}
\begin{flushright}
\texttt{\footnotesize CERN-PH-TH/2009-217}
\end{flushright}
\vfill
Nuclear physics from strong coupling QCD
}

\ShortTitle{Nuclear physics at strong coupling}

\author{Michael Fromm\\
	Institute for Theoretical Physics, ETH Z\"urich, CH-8093 Z\"urich, Switzerland\\ 
	E-mail: \email{fromm@phys.ethz.ch}} 
	
\author{Philippe de Forcrand\\ 
Institute for Theoretical Physics, ETH Z\"urich, CH-8093 Z\"urich, Switzerland\\
and\\
CERN, Physics Department, TH Unit, CH-1211 Geneva 23, Switzerland\\ 
E-mail: \email{forcrand@phys.ethz.ch}} 

\abstract{The strong coupling limit ($\beta_{\rm gauge} = 0$) of QCD offers 
a number of remarkable research possibilities, of course at the price of large 
lattice artifacts. Here, we determine the complete phase diagram as a function
of temperature $T$ and baryon chemical potential $\mu_B$, for one flavor of 
staggered fermions in the chiral limit, with emphasis on the determination of 
a tricritical point and on the $T\approx 0$ transition to nuclear matter. 
The latter is known to happen for $\mu_B$ substantially below the baryon 
mass, indicating strong nuclear interactions in QCD at infinite gauge coupling. 
This leads us to studying the properties of nuclear matter from first principles.
We determine the nucleon-nucleon potential in the strong coupling limit, as 
well as masses $m_A$ of nuclei as a function of their atomic number $A$. 
Finally, we clarify the origin of nuclear interactions at strong coupling, 
which turns out to be a steric effect.}

\FullConference{The XXVII International Symposium on Lattice Field Theory - LAT2009\\
		 July 26-31 2009\\
		 Peking University, Beijing, China}

\begin{document}
\section{Model and motivation}
We study lattice QCD with one species of staggered fermions at infinite gauge coupling~\cite{Karsch88}. The partition function
\be
	Z (m_q, \mu) = \int\mathcal{D}U\mathcal{D}\bar\chi\mathcal{D}\chi ~ \mathrm{e}^{S_\mathrm{F}} ,
\label{eq:Z}
\ee
\nopagebreak
depends on the staggered quark mass $m_q$ and the baryon chemical potential $\mu_B$.  Since $\beta_{\mathrm{gauge}} = 0$, we only have the fermionic action $S_{\mathrm{F}}$ with
\be
	S_\mathrm{F} = \!\!\!\! \sum_{x,\nu=1,4} \!\!\! \eta_{x,\hat{\nu}}\bar\chi_x\left[U_{x,\hat\nu}\chi_{x+\hat\nu} - U^{\dagger}_{x-\hat\nu,\hat\nu}\chi_{x-\hat\nu}\right] + 2m_q\sum_x\bar\chi_x\chi_x\label{sc_action}
\ee
where $\eta_{x,\hat\nu} \!=\! (-1)^{\sum_{\rho<\nu}x_\rho}$, $\eta_{x,\hat1} \!=\! 1$. The quark chemical potential $\mu$ and an anisotropy $\gamma$ are 
introduced by multiplying the time-like gauge links $U_{x,\pm\hat{4}}$ by 
$\gamma\exp(\pm a\mu/\gamma)$ in the forward and backward directions, respectively. 
The anisotropy allows to vary the temperature continuously, via $T \!=\! a^{-1} \gamma^2/N_\tau$ at infinite coupling~\cite{Boyd91, Damgaard}.

When $m_q = 0$, our model is invariant under global $U(1)_A\times U(1)_B$ transformations
\be
\chi(x) \to e^{i(\varepsilon(x)\theta_A+\theta_B )}\chi(x),~
\bar\chi(x) \to \bar\chi(x) e^{i(\varepsilon(x)\theta_A-\theta_B)} 
~\forall ~x\,,
\label{eq:chiral}
\ee
where $\varepsilon(x) \!=\! (-1)^{\sum_4 x_\nu}$.  While $U(1)_B$ is responsible for baryon number conservation and remains unbroken, $U(1)_A$ represents the chiral symmetry of the model and can break spontaneously.

Owing to the absence of the gauge action in Eq.~(\ref{sc_action}), the link integration factorizes in Eq.~(\ref{eq:Z}) and can be done analytically~\cite{Rossi1984}. The degrees of freedom are now mesons and baryons. Carrying out the Grassmann integration, the partition function Eq.~(\ref{eq:Z}) becomes for $m_q=0$ that of a dimer-loop model~\cite{Karsch88}:
\be
        Z(m_q = 0,\mu) = \displaystyle\sum_{\left\{n_{x,\hat\nu},C\right\}}
        \prod_{x,\hat\nu}\gamma^{2\delta_{\nu,4} n_{x,\hat 4}}\frac{(3-n_{x,\hat\nu})!}{n_{x,\hat\nu}!}
        ~ \prod_{C}w(C)\,,
\label{eq:Z_loop}
\ee
with the constraint that mesonic links with occupation number $n_{x,\hat\nu} = 0,..,3$ attached to a site $x$ satisfy $\sum_{\pm\mu} n_{x,\pm\hat\mu} = 3$. 
Alternatively, a site can be traversed by a self-avoiding baryon loop $C$. 
The weight of such a loop $C$ is given by $w(C) \!=\! \rho(C) \gamma^{3N_{\hat{4}}(C)}\exp{\left(3k\mu a N_\tau/\gamma\right)}$. Here, $N_{\hat{4}}(C)$ is the number of links on the loop in the time direction, $k$ is its winding number in this direction 
and $\rho(C)\!=\! \pm 1$ is a geometry-dependent sign, which can be negative even when $\mu\!=\! 0$. 
Karsch and M\"utter~\cite{Karsch88} removed the sign problem present already 
at $\mu = 0$ by analytically resumming pairs of configurations. 
When $\mu >0$ the remaining sign problem is mild (see next Section), allowing us
to simulate large enough lattices that the phase diagram of the model can be
determined reliably.

There exists a plethora of mean-field results dating back to the early days of lattice QCD~\cite{MF_pioneer} and continuing up to now~\cite{Kawamoto2007, Nishida2003}, with the inclusion of NLO and NNLO corrections to the infinite coupling limit~\cite{Miura2009_NNLO}, and statements about a quarkyonic phase~\cite{Miura2008_quarkyonic}. These approximate predictions should be checked against numerical simulations using an exact algorithm.

Moreover, lattice QCD should provide the means to study nuclear physics from first principles. In lattice QCD at \emph{weak coupling}, properties of single hadrons~\cite{Durr2008}, nuclear scattering lengths and potentials~\cite{Beane_Savage, Ishii:2008, Gardestig:2009} or two-and three baryon systems~\cite{Beane:2009} are being studied. Going to higher nuclear matter density is mostly hindered by a severe sign problem. Contrary to that, the strong coupling limit offers not only the possibility to study nuclear matter at higher density thanks to the mild sign problem, but also provides an intuitive understanding, owing to the point-like nature of baryons and pions in this limit. 
The absence of earlier numerical studies of the strong coupling version of 
nuclear physics is mostly due to algorithmic issues present in the early 
studies~\cite{Karsch88, Aloisio1999}. Decisive progress occurred with the 
introduction of the worm algorithm~\cite{Prokof'ev2001}, which has been adapted 
for strong coupling $SU(2)$ and $U(3)$ lattice theories in~\cite{Shailesh}, 
enabling efficient Monte Carlo sampling even in the chiral limit $m_q \!\to\! 0$.
We extend this approach to $SU(3)$.

\begin{figure}[t]
	\centerline{
	\subfigure[]{\includegraphics*[scale=0.4]{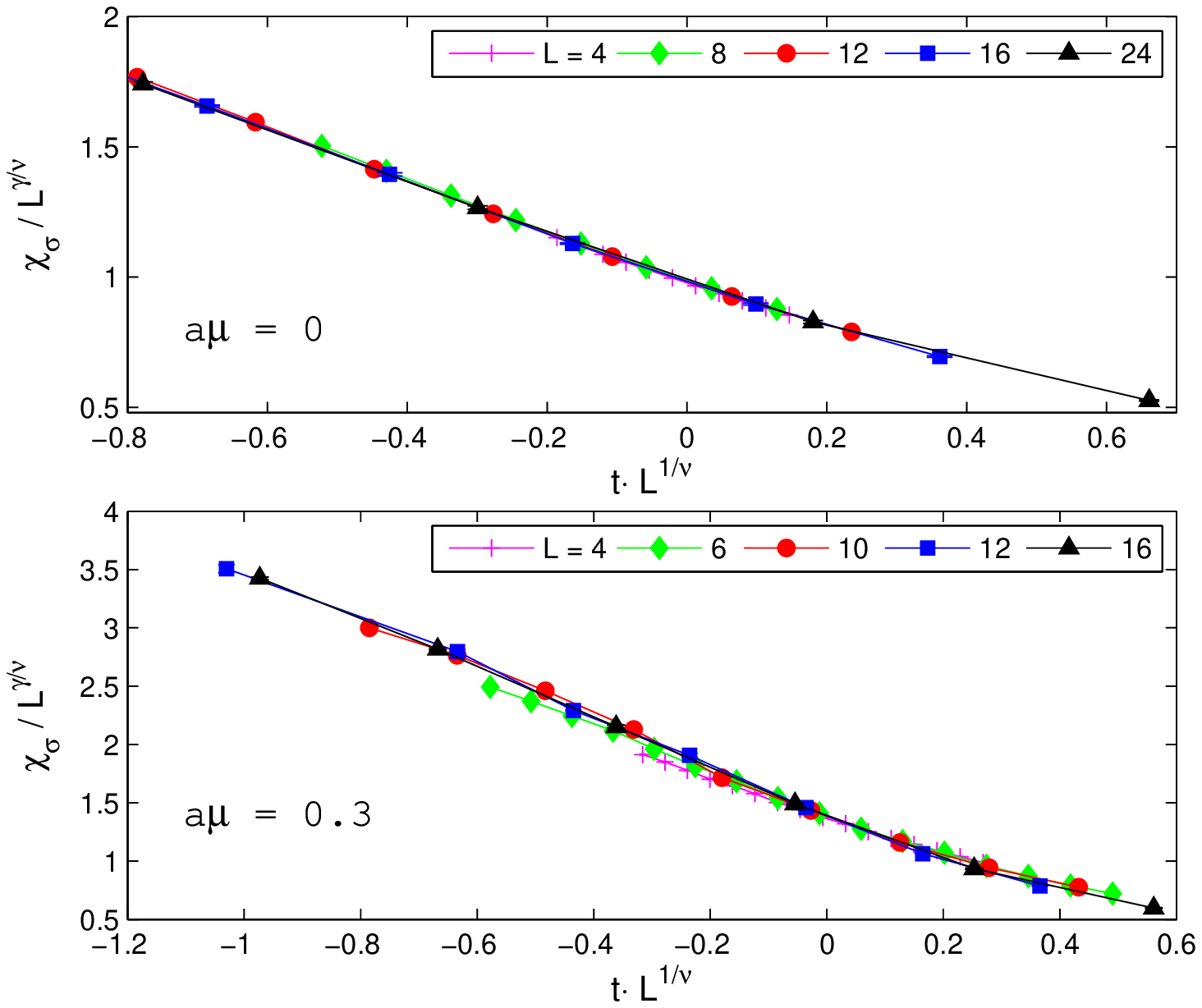}
	\label{fig:chi_collapse}}
	\subfigure[]{\includegraphics*[scale=0.4]{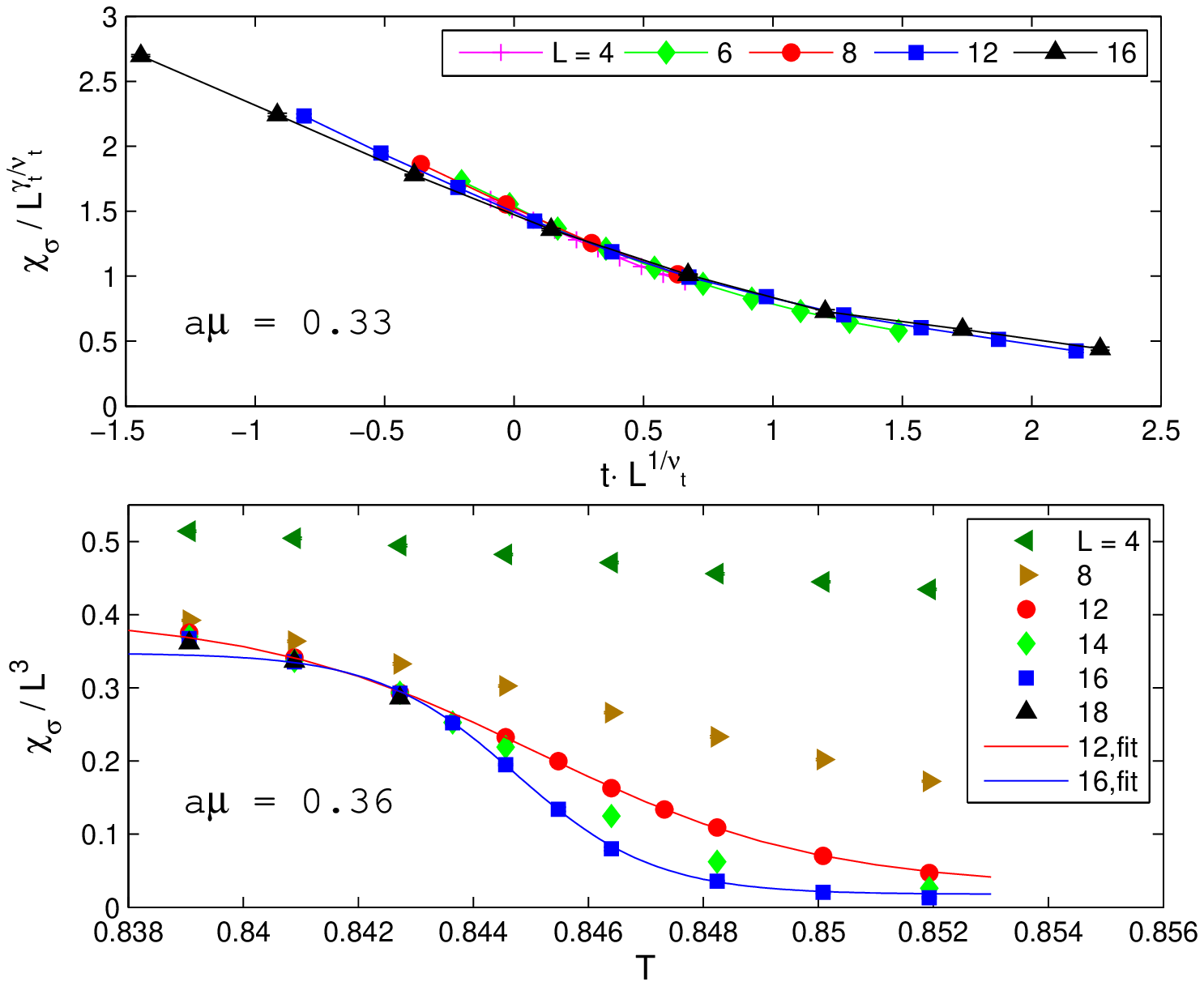}
	\label{fig:chi_collapse_tcp_1st}}}
	\caption{Finite-size scaling of the chiral susceptibility near the phase
transition, with ($a$) $d=3,~O(2)$ exponents at  $a\mu = 0$ and $a\mu = 0.3$, 
and ($b$,top) tricritical (mean-field) exponents at $a\mu = 0.33$. 
Lines are drawn to guide the eye. 
For $a\mu = 0.36$ ($b$, bottom), we instead rescale the $y$-axis with 
$L^{-\gamma/\nu} = L^{-3}$ and observe that $\chi_{\sigma}/L^3$ goes to a 
constant for $T<T_c(\mu)$. The solid lines are fits derived from the first-order
ansatz Eq.~(\protect\ref{eq:borgs}). 
All results have been obtained for $N_\tau = 4$. 
Errors are of the size of the symbols.}
\end{figure}
\section{Phase diagram}
Since we consider here the case of a massless quark, the chiral symmetry $U(1)_A$ Eq.~(\ref{eq:chiral}) is exact but 
spontaneously broken at small $(T,\mu)$, with order parameter $\langle \bar\psi \psi \rangle$.
When $\mu\!=\!0$, a mean-field analysis predicts symmetry restoration at $a T_c\!=\!5/3$, whereas the Monte Carlo 
study of \cite{Boyd91} on $N_\tau\!=\!4$ lattices, extrapolated to $m_q\!=\!0$, finds $a T_c \!=\! 1.41(3)$.
In order to determine $T_c$ we performed numerical simulations at $a \mu=0, a m_q = 0$. Our main observable is the chiral susceptibility 
\be
\chi_\sigma = \frac{1}{V}\left.\frac{\partial^2}{\partial m_q^2}\log{Z}\right|_{m_q=0}= \langle\sum_x\bar\psi\psi(x)\bar\psi\psi(0)\rangle\, ,
\ee
where we used the fact that in a finite volume the order parameter  $\langle \bar\psi \psi \rangle$  vanishes identically. Correspondingly, $\chi_\sigma$ will not show a peak at the transition but finite size scaling (FSS) still applies,
so that for a system of size $L^3\times N_\tau$ at a ``reduced temperature''
$t = 1 - T/T_c$ (or $1 - \mu/\mu_c$):
\be
\chi_\sigma \sim L^{\gamma/\nu}\tilde{\chi}(t L^{1/\nu}) = L^{\gamma/\nu}\left(a_0+\mathcal{O}(t L^{1/\nu})\right )\,.
\label{eq:chi_scaling}
\ee
In principle, this allows us to determine the exponents $\gamma$ and $\nu$, 
which should be those of the $d=3,~O(2)$ universality class~\cite{Campostr2000}.
In practice however, the numerical values are close to those of the $Z(2)$ 
universality class and we point to~\cite{Chandrasekharan2003} for a more 
advanced treatment in the case of U(3). Here, we simply assume $d=3,~O(2)$ 
exponents and show that $\chi_\sigma$ collapses on a universal curve when rescaled 
according to Eq.~(\ref{eq:chi_scaling}). For example, 
Fig.~\ref{fig:chi_collapse} (top) shows the collapse for $\mu=0, N_\tau=4$
and several values of $L$, with $aT_c$ fixed to $1.402$ which is the crossing
point of $\chi_\sigma/L^{\gamma/\nu}$ when plotted as a function of $T$. 
Using this strategy, we find $a T_c \!=\! 1.319(2),1.402(3),1.417(3)$, 
respectively, for $N_\tau\!=\!2, 4, 6$, indicating an $N_\tau \!\to\! \infty$ 
limit about $15\%$ smaller than the mean-field prediction. 

\begin{figure}[t]
	\centerline{
	\subfigure[]{\includegraphics*[scale=0.278]{pd.eps}
	\label{fig:phase_diagram}}
	\subfigure[]{\includegraphics*[scale=0.41]{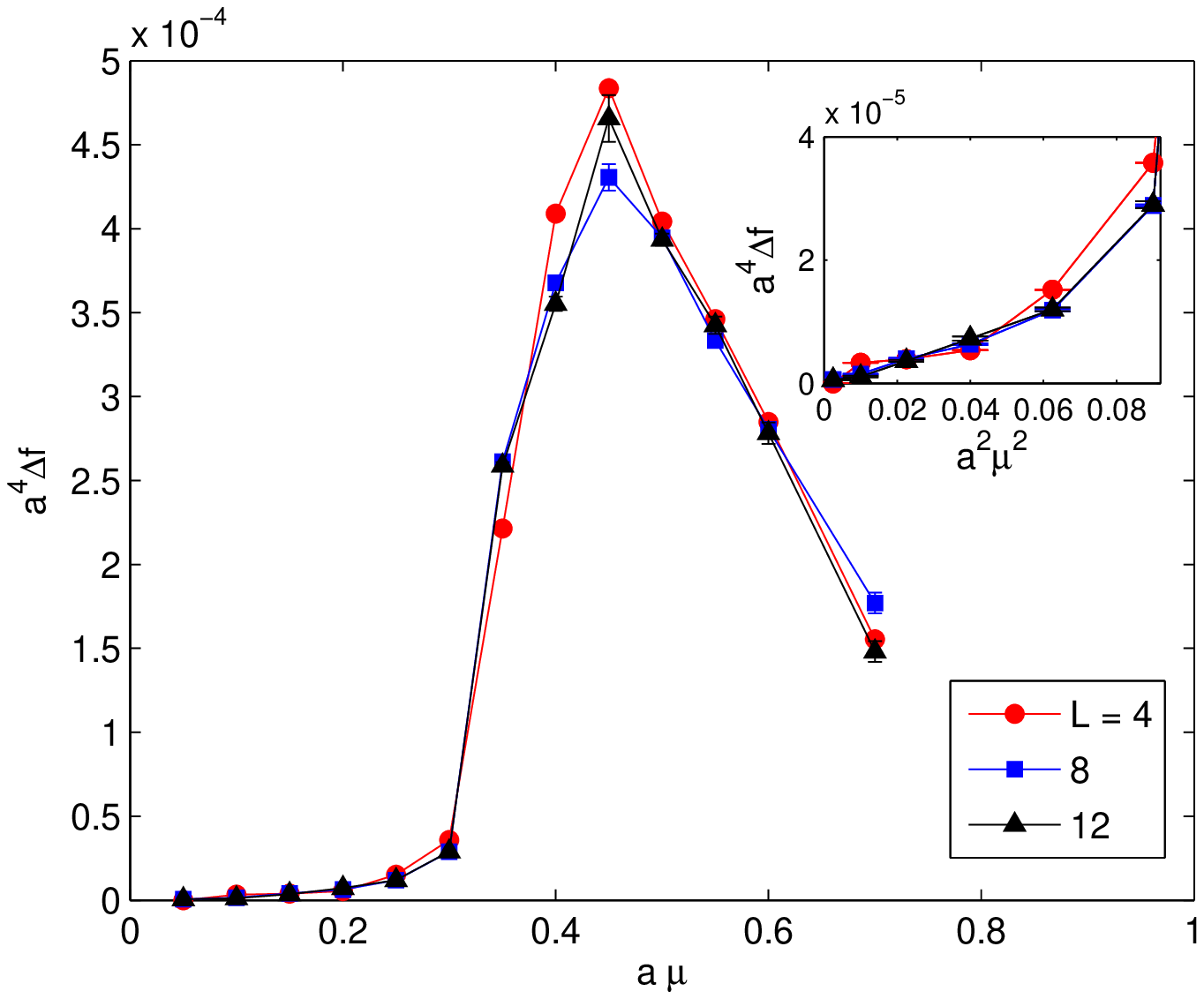}
	\label{fig:av_sign}}}
	\caption{($a$)~$(\mu,T)$ phase diagram of 1-flavor strong coupling QCD 
with massless staggered fermions ($N_\tau=4$).
($b$)~$a^4 \Delta f = -\gamma^2\log{(\langle{\rm sign}\rangle)}/(L^3N_\tau)$ 
versus $a\mu$ at $T = T_{\mathrm{TCP}}$. Lines are drawn to guide the eye.}
	
\end{figure}
For $\mu >0$ the model has a sign problem. We have measured the ``average sign''
\be
\langle{\rm sign}\rangle = \frac{Z}{Z_{\|}} =  \exp{(-\Delta fV/T)} = \exp({-a^4\Delta fL^3N_\tau/\gamma^2})\,,
\ee
where $Z_\|$ corresponds to the partition function $Z$ of Eq.~(\ref{eq:Z_loop}),
but taking the absolute value of the weights. The free energy density 
difference $(a^4 \Delta f)$ is a measure of the severity of the sign problem.
Fig.~\ref{fig:av_sign} shows $(a^4\Delta f)$ as a function of $a\mu$, for
 several $L^3\times 4$ lattices at $T = 0.937 a^{-1} \approx T_{\rm TCP}$ (see below), using the 
analytic resummation prescription of Karsch and M\"utter~\cite{Karsch88} which
removes the sign problem at $\mu=0$. $\Delta f$ is nicely volume-independent,
vanishes at $\mu=0$, starts $\propto \mu^2$ (see inset), and peaks slightly past
the phase transition. Note the very small magnitude ${\cal O}(10^{-4})$ 
-- compared to $(a^4 \Delta f) \sim {\cal O}(1)$ expected when using the standard
approach of integrating over the fermions first -- which allows us to simulate
$16^3\times 4$ lattices with $\langle {\rm sign}\rangle \ge 0.1$. 

\begin{figure}[t]
	 \centerline{	\subfigure[]{\includegraphics*[scale=0.13]{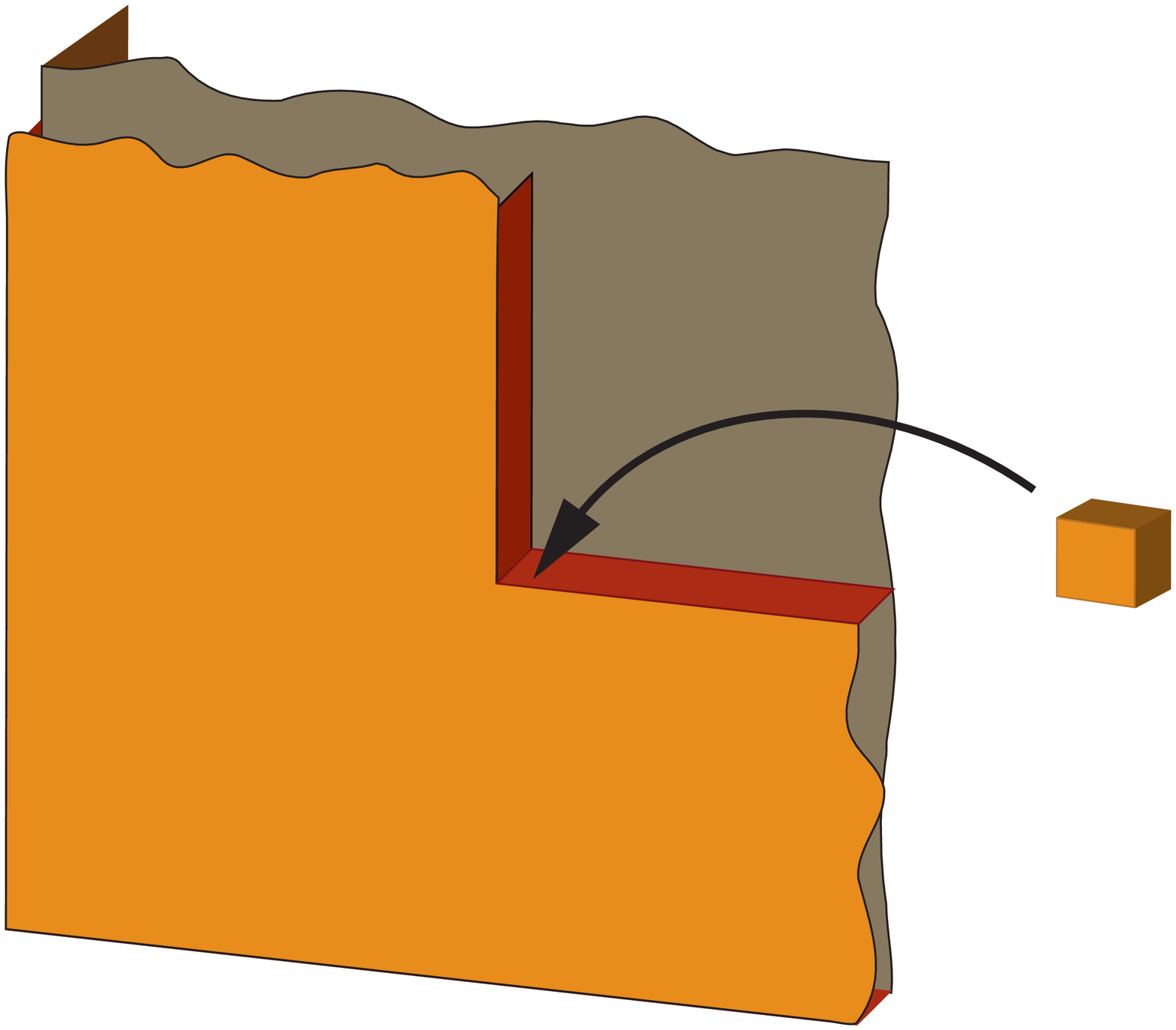}
	\label{fig:domain_wall}}
	\subfigure[]{\includegraphics*[scale=0.13]{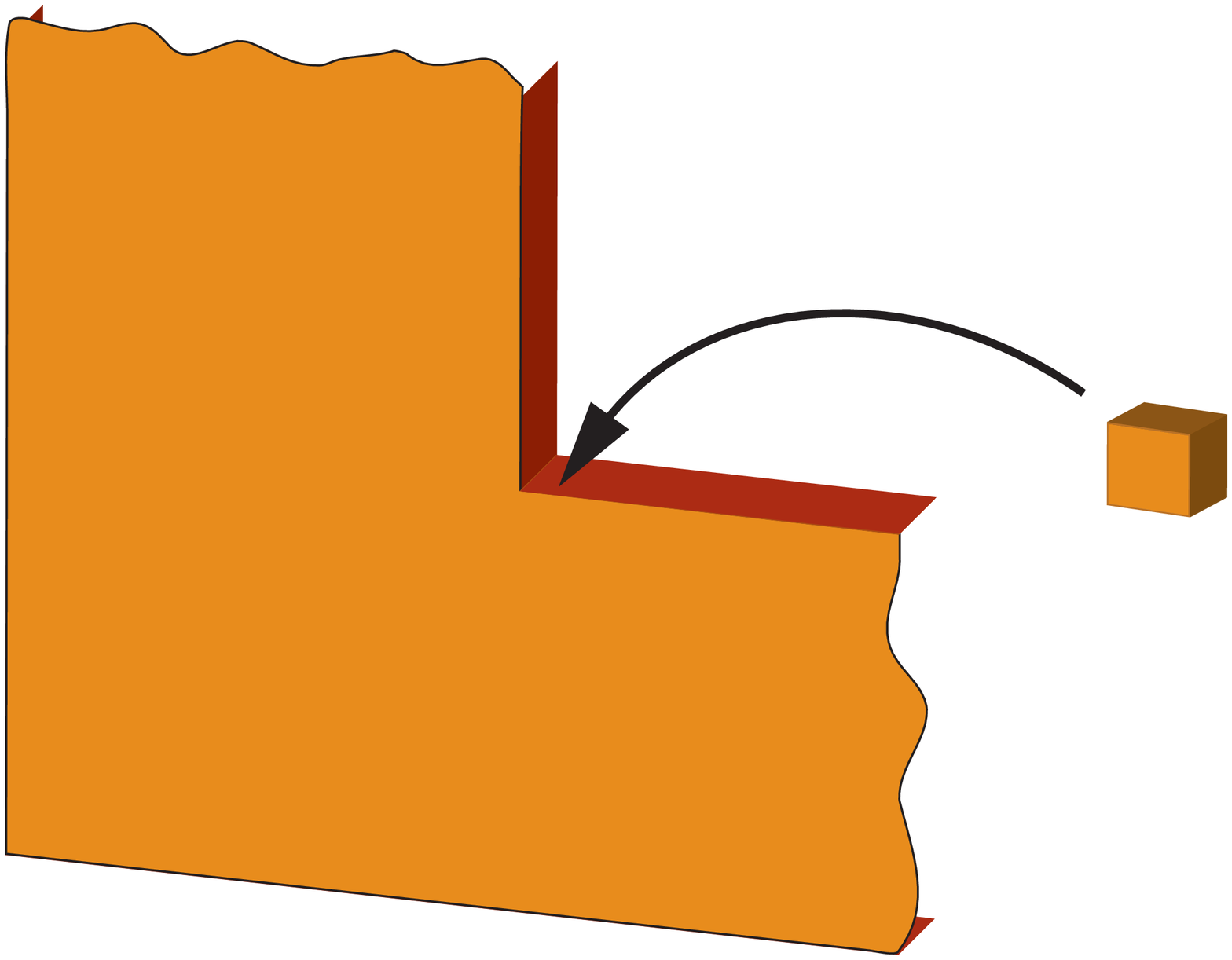}
	\label{fig:domain_wall_only}}
	 \subfigure[]{\includegraphics*[scale=0.4]{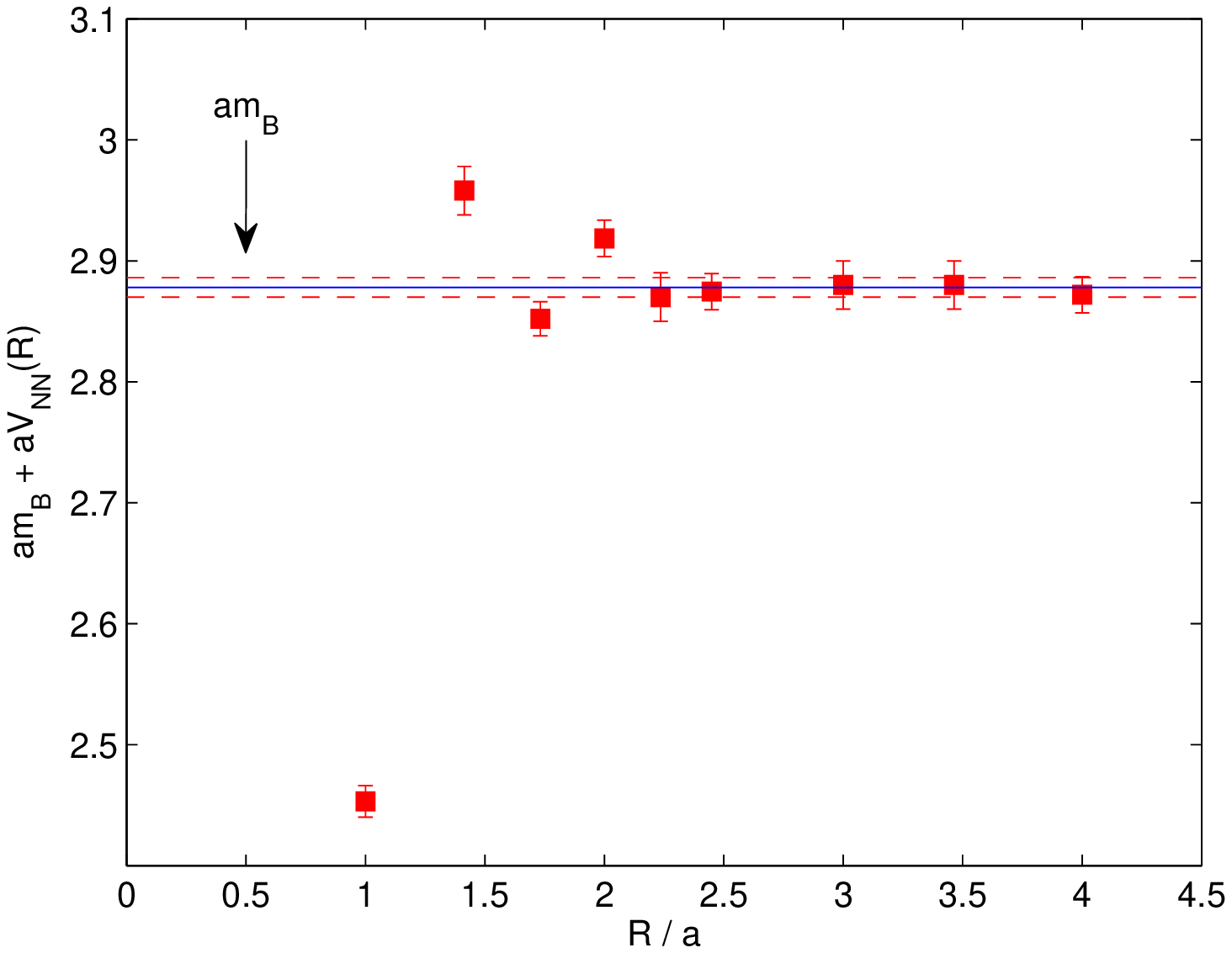}
	 \label{fig:V_NN}}}
	\caption{($a$) Adding a baryon to grow an additional layer of bulk 
nuclear matter: each new baryon binds to 3 nearest-neighbors.
($b$) Building a first layer of nuclear matter inside the hadron gas, 
thus creating two interfaces: each new baryon binds to 2 nearest-neighbors.
($c$) Energy of a second static baryon at distance $R$ from the first, 
{\it i.e.}, $(m_B + V_{\rm NN}(R))$, where $V_{\rm NN}(R)$ is the nuclear 
interaction potential. The horizontal band indicates the mass of an isolated 
baryon and corresponds to $V_{\rm NN}=0$. At $R\!=\!0$ the potential is 
infinitely repulsive.}
\end{figure}

For the available volumes, we may then follow the critical line as $a\mu$ 
increases, monitoring the collapse of $\chi_\sigma$ using the appropriate 
critical exponents. Expectations are that the second order $O(2)$ transition 
will turn first order at a tricritical point (TCP) for some nonzero $\mu$. 
From Fig.~\ref{fig:chi_collapse} (bottom), we see that for $a\mu = 0.3$, 
$\chi_\sigma$ still obeys $O(2)$ scaling behavior. With a slight increase to
$a\mu = 0.33$  Fig.~\ref{fig:chi_collapse_tcp_1st} (top), a satisfactory 
collapse requires tricritial (mean-field) exponents $\gamma_t = 1, \nu_t = 1/2$.
Under a further small increase to $a\mu  = 0.36$ no such finite-size scaling
collapse can be achieved.
Instead, the transition is first-order:
$(i)$ the distribution of the baryon density shows two peaks, whose areas 
become equal at $a T_c = 0.844(1)$;
$(ii)$ for $T<T_c$, $\chi_{\sigma}/L^3$ becomes $L$-independent (see 
Fig.~\ref{fig:chi_collapse_tcp_1st} (bottom));
$(iii)$ on either side of $T_c$, $\chi_\sigma$ is well described by differentiating the
two-phase Borgs-Kotecky ansatz~\cite{Borgs}
\be
Z(T) = \exp(-\frac{V}{T} f_1(T)) + c \exp(-\frac{V}{T} f_2(T))
%\chi_\sigma(T) = \frac{a(T)+b(T)L^3\exp{(-\delta f(T-T_c)L^3)}}{1+c\exp{(-\delta f (T-T_c)L^3)}}\,.
\label{eq:borgs}
\ee
with $f_{1,2}(T) = f_0 \pm \alpha (T-T_c)$, from which the solid curves can
be obtained. Thus, from the available data we conservatively conclude that 
$(a \mu_{\rm TCP},a T_{\rm TCP}) \!=\! (0.33(3), 0.94(7))$. This should be 
compared with the analytic prediction $(0.577,0.866)$ of~\cite{Nishida2003}. 
The rather large difference in $\mu$ underlines the ${\cal O}(1/d)$ accuracy of a mean-field treatment, and justifies a posteriori our Monte Carlo study.
In Fig.~\ref{fig:phase_diagram} we summarize our findings for the phase diagram. In spite of its resemblance to the expected deconfinement transition in massless $N_f\!=\!2$ QCD, 
here the two phases are both confining, with point-like mesons and baryons, and so the phase transition is to dense, chirally symmetric
nuclear matter. At $T\!=\!0$ the baryon density jumps from 0 to 1, a saturation value caused by the self-avoiding
nature of the baryon loops, which itself originates from their fermion content. Using the baryon mass to fix the lattice spacing, this 
represents about 4 nucleons per fm$^3$, around 25 times the real-world value. An intriguing feature of this $T\!=\!0$ transition - and an important motivation for this study - is the value of $\mu_B^{\rm critical}$, which both mean-field~\cite{MF_pioneer} and an early Monte Carlo study~\cite{Karsch88} find much smaller than the naive threshold value $m_B$. 
However, the ergodicity of the simulations of \cite{Karsch88} was questioned in~\cite{Aloisio1999}, which was found to be justified in~\cite{Fromm:2008}.
This motivated us to redetermine $\mu_B^{\rm crit}(T\!=\!0)$ using an improved method inspired by the ``snake'' algorithm~\cite{deForcrand2000}: When two phases coexist,
the free energy $\Delta F/T$ necessary to increase by a ``slice'' $L\times L \times a$ the volume occupied by dense nuclear matter can be decomposed into $L^2$
elementary contributions, looking generically like Fig.~\ref{fig:domain_wall}, where one additional static baryon is 
attached to 3 neighbors.
We measured the free energy $\Delta F/T$ of this elementary increment on 
a large $8^3\times16$ lattice, and obtained $a \Delta F \!=\! a \mu_B^{\rm crit} \!=\! 1.78(1)$, rather close to
both mean-field predictions~\cite{MF_pioneer} and Monte Carlo extrapolations~\cite{Karsch88}. This we compare to $a m_B$ which corresponds to the difference in free energy measured at $\mu=0, T\approx 0$ by extending a static baryon world line. 
We find $a m_B =2.88(1)$, consistent with HMC simulations~\cite{Kim}, and again in agreement with mean-field~\cite{MF_pioneer} and large-$N_c$~\cite{Martin1983} predictions but indeed much larger than $a \mu_B^{\rm crit}$.  As already recognized in \cite{Bilic:1991}, the reason that $\mu_B^{\rm crit} < m_B$ must then be the presence of a strong nuclear attraction, to which we now turn.

\section{Nuclear Physics at infinite gauge coupling}
Since our baryons are point-like, there is no conceptual difficulty in defining the nuclear potential
$V_{\rm NN}(R)$, unlike in the real world~\cite{Beane2008}. We measured $V_{\rm NN}$ using again the ``snake''
algorithm, this time extending little by little in Euclidean time the worldline of a second baryon at distance
$R$ from the first. The result is shown in Fig.~\ref{fig:V_NN}. Aside from the hard-core repulsion, there is indeed
a strong nearest-neighbor attraction, a slight repulsion at distance $a\sqrt{2}$, and almost no 
interaction beyond that distance. $V_{\rm NN}$ is similar qualitatively to what is expected in the real
world, with competition between attractive $\sigma$ exchange and repulsive $\omega$ exchange.
The depth of the minimum $\sim\!120$~MeV and the corresponding distance $\sim\! 0.6$~fm are quantitatively plausible~\cite{Kapusta}. This nearest-neighbor attraction also explains a posteriori the value of $\mu_B^{\rm crit}$: each baryon
added to the dense phase binds with 3 nearest neighbors, which reduces the increase in free energy from $a m_B$ to only
$a(m_B + 3 V_{\rm NN}(a)) \approx 1.7$, consistent with $a\mu_B^{\rm crit}$.

Similarly, we can predict the $T\!=\!0$ surface tension
of nuclear matter: in a periodic cubic box, when building a first ``slice'' of
nuclear matter with two interfaces in the dilute phase, each new baryon binds with
only 2 nearest-neighbors (Fig.~\ref{fig:domain_wall_only}) instead of 3 in the bulk (Fig.~\ref{fig:domain_wall}), thus increasing its free energy by $|V_{\rm NN}(a)|$ for an increase of $2 a^2$ in the interface area, yielding $\sigma \!\approx\! \frac{a^{-2}}{2} |V_{\rm NN}(a)|$.

\begin{figure}[t!]
	\centerline{
		\subfigure[]{\includegraphics*[scale=0.4]{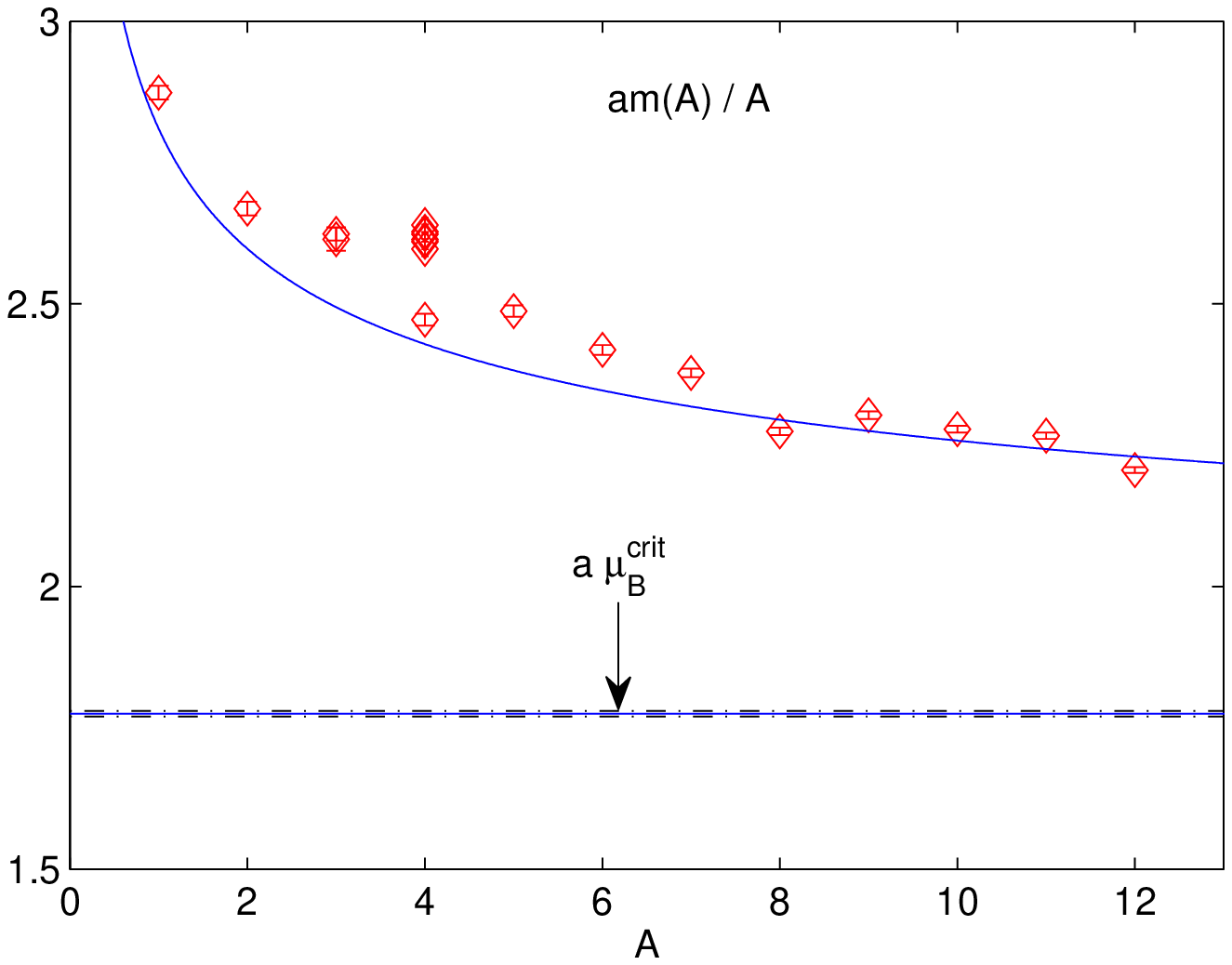}
		\label{fig:masses}}
		\subfigure[]{\includegraphics*[scale=0.15]{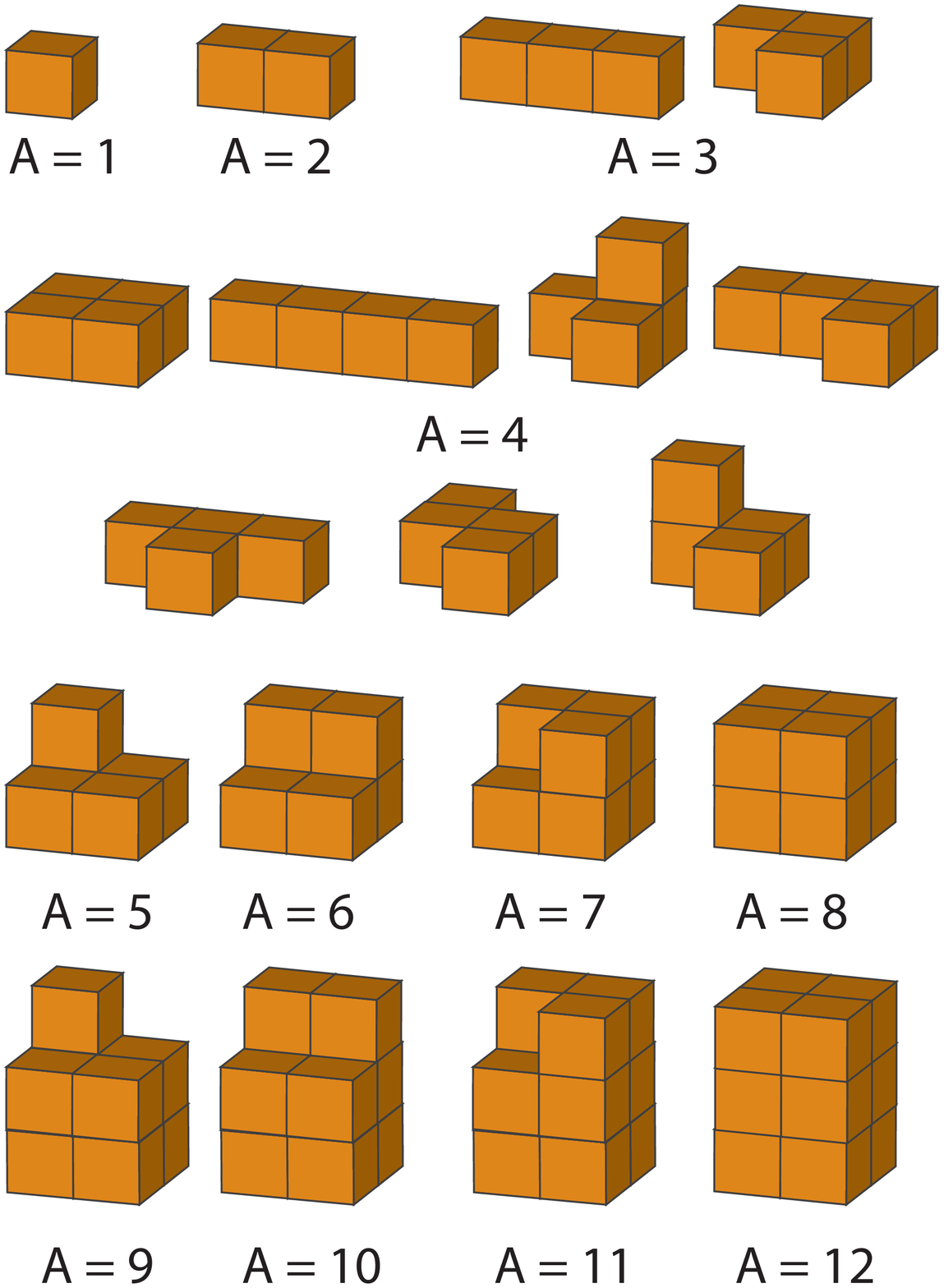}
		\label{fig:geometries}}
		\subfigure[]{\includegraphics*[scale=0.40]{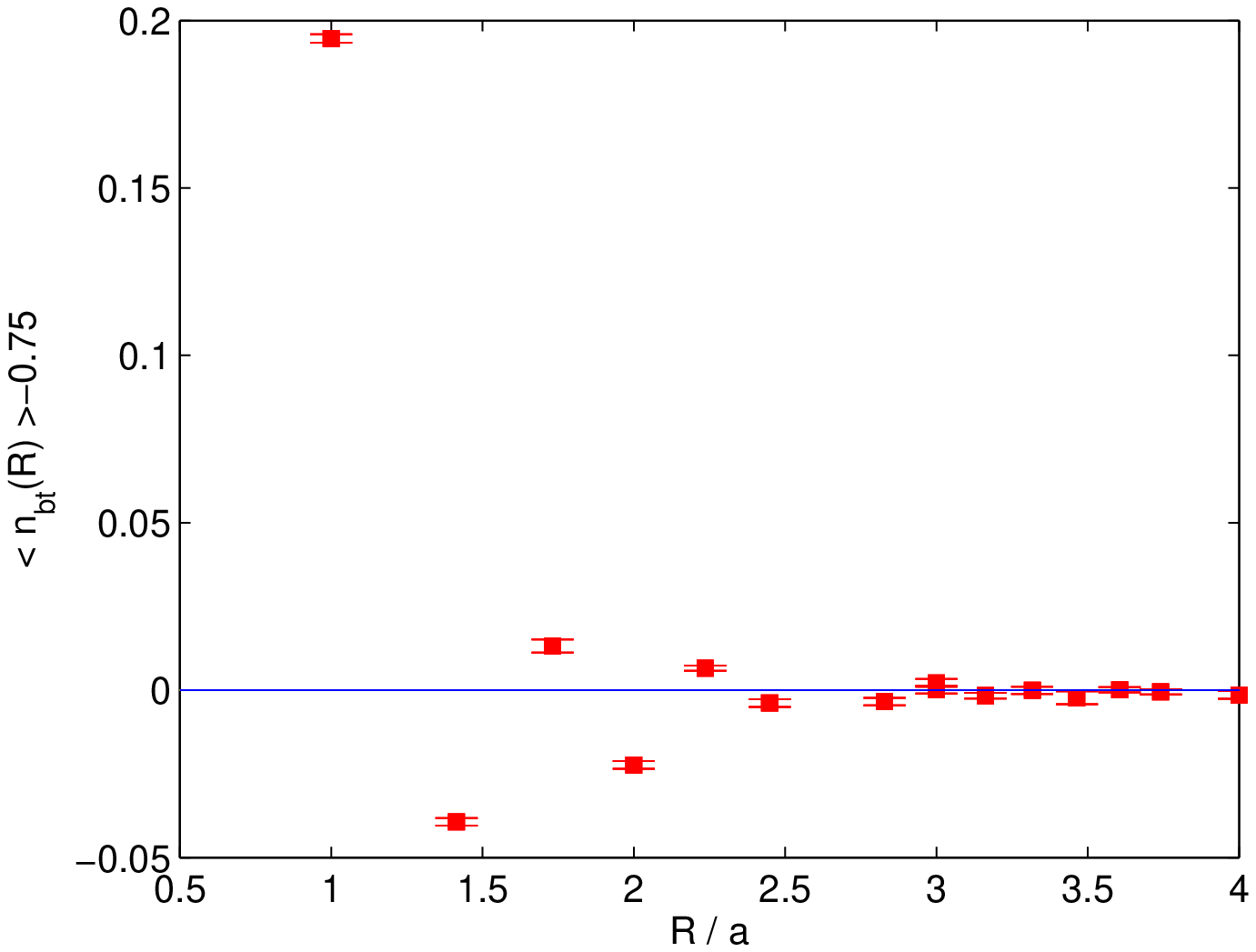}
		\label{fig:n_bt}}}
	\caption{($a$) Mass per nucleon of $A\!=\!1,..,12$ nuclei. For 
$A\!=\!3,4$ all possible geometric isomers are included. The solid line shows 
the parameter-free Bethe-Weizs\"acker Eq.~(\protect\ref{eq:Bethe_Weiz}), with 
the surface tension $\sigma$ set to $\frac{a^{-2}}{2} |V_{\rm NN}(a)|$.
($b$) Corresponding nuclear geometries in order of increasing mass.
($c$) Energy density of the pion cloud as a function of the Euclidean distance 
to a static baryon.}
\end{figure}

This large interface tension has an impact on the stability of nuclei of various sizes and shapes:
for a given atomic number $A$, those with a shape close to a sphere (or a cube) will have a smaller mass.
Using the same variant of the ``snake'' algorithm, we have added baryons, one by one, to form such nuclei
while measuring the successive increments in free energy.
For $A\!=\!2$ our ``deuteron'' binding energy is about
120 MeV: the real-world binding energy of $\sim\! 2$~MeV results from delicate
cancellations which do not occur in our 1-flavor model, and the binding
energy remains of the same magnitude as the depth of $V_{\rm NN}$.
For larger $A$, the resulting Fig.~\ref{fig:masses} does indeed show increased stability for
nuclei having square ($A\!=\!4$), cubic ($A\!=\!8$) or parallelepipedic ($A\!=\!12$) shapes. Other ``isomers'' with different
shapes, studied exhaustively for $A\!=\!4$ and sketched Fig.~\ref{fig:geometries},
have clearly larger masses. Moreover, the average mass per nucleon
is well described by the first two (bulk and surface tension) terms of the Weizs\"acker phenomenological formula:
\be
m(A)/A = \mu_B^{\rm crit} + (36\pi)^{1/3} a^2 \sigma A^{-1/3},
\label{eq:Bethe_Weiz}
\ee
where $\sigma$ is set equal to $\frac{a^{-2}}{2} |V_{\rm NN}(a)|$ in
the Figure. The next higher-order terms in this formula come from isospin and Coulomb forces, which are both absent
in our model.

\section{Summary and Remarks}
An interesting aspect of our study is the {\em origin} of the nuclear interaction.
The nucleons are point-like and self-avoiding, so that only the hard-core repulsion is explicit.
There is no pion exchange. In a way reminiscent of the Casimir effect between two neutral plates,
the interaction proceeds by the rearrangement of the pion bath caused by the
excluded volume of the nucleon.
This rearrangement is visible Fig.~\ref{fig:n_bt}
for one nucleon: at a neighboring site, the three pion lines attached
to each site have fewer options and orient more often along the Euclidean time, which increases the pion
energy. In fact, the nucleon mass $a m_B \!\approx\! 2.88$ can be decomposed into a bare mass $3 - 3/4 \!=\! 2.25$,
which is the energy increase ``inside'' the nucleon and can be assigned to
the three valence quarks, and an energy increase $\approx\! 0.63$ in the
surrounding pion ``cloud''. When two nucleons are next to each other, the latter increase is limited to 10 nearest-neighbors
instead of $2\!\times\! 6$, which explains the attraction between them (in sign and roughly in magnitude).
This excluded volume or ``steric'' effect is thus the origin of the nuclear potential, and ultimately of
nuclear stability, in our model. In real QCD, the pion density is not constrained as in Eq.~(\ref{eq:Z_loop}).
Nevertheless, it is going to be high at temperatures $T\!\sim\! m_\pi$~\cite{Gerber1988} and one should expect
the same steric effect to enhance nuclear attraction at such temperatures. 

To summarize, in a crude model of QCD, 1-flavor massless lattice staggered fermions at strong coupling $\beta_{\rm gauge}\!=\!0$, we have
been able to obtain the complete phase diagram and derive the strong coupling version of nuclear interactions and nuclear masses  
from first principles, uncovering a simple, but universal, steric origin of the nuclear interaction.
This model can be improved in many ways.
One simple modification consists of giving a non-zero mass to
the quarks: the nuclear interaction will weaken as the pion mass is increased,
in a way which can be compared with effective field theories.
Less simple but feasible improvements include introducing isospin with
a second quark flavor, and measuring the ${\cal O}(\beta)$ correction
as done analytically in \cite{Faldt:1985,Bilic_Karsch:1991, Miura2009_NNLO}. 
These will bring our model much closer to real QCD.

\emph{Acknowledgments}. The work of M.F. was supported by ETH Research Grant TH-07 07-2.

\end{document}